CrossMark

# Interior point search for nonparametric image segmentation

**Sinan Onal[1]** · **Xin Chen[1]** · **Madagedara Maduka Balasooriya[2]**

**Abstract** Precise object boundary detection for automatic image segmentation is critical for image analysis, including that used in computer-aided diagnosis. However, such detection traditionally uses active contour or snake models requiring accurate initialization and parameter optimization. Identifying optimal parameter values requires time-consuming multiple runs and provides results that vary by user expertise, limiting the use of these models in high-throughput or real-time situations. Thus, we developed a nonparametric snake model using an interior point search method applied in iterations to find and improve the set of snake points forming the edge of a shape. At each iteration, one or more snake points are replaced by others in the edge map. We validated the model using binary and continuous edge images of single and multiple objects, and noisy and real images, comparing the results to those obtained using traditional snake models. The proposed model not only provides better results on all image types tested but is more robust than traditional snake models. Unlike traditional snake models, the proposed model requires no user interaction for initializing snakes and no pre-processing of noisy images. Thus, our method offers robust automatic image segmentation that is simpler to use and less time-consuming than traditional snake models.

**Keywords** Active contour model · Digital image processing · Interior point search · Segmentation · Continuous edge map

✉ Sinan Onal
sonal@siue.edu

[1] Department of Mechanical and Industrial Engineering, Southern Illinois University Edwardsville, Box 1805, Edwardsville, IL 62026, USA

[2] Department of Mathematics and Statistics, Southern Illinois University, Edwardsville, IL 62026, USA

## 1 Introduction

Subdividing an image into its constituent regions, called image segmentation, is a crucial task in image processing. The success or failure of image subdivision is often a direct consequence of the success or failure of the segmentation. Myriad segmentation techniques have been proposed in the literature involving the detection, recognition, and measurement of objects in digital images, including industrial inspection [1,2], optical character recognition [3], object tracing in a sequence of images [4], and anatomical structure detection, such as the detection and measurement of bone, soft tissue, or tumor, in medical images [5,6]. Overall, two image segmentation approaches exist: algorithms that are based on models and those that are model-free.

Model-free segmentation methods apply statistical tools, such as histogram, mean, variance, and entropy, to local image properties. Among the most frequently used model-free approaches, including thresholding, image feature space clustering, and region- and boundary-based methods [7–14].

Segmentation approaches using models can be categorized as either training or deformable based.

One of the main deformable model-based approaches is active contour or "snake" segmentation [15]. An active contour detects specified properties of an image and dynamically fits to the edges of a structure by minimizing an energy function. Although successful applications have been reported for the segmentation of objects [16–21], the active contour approach has several difficulties. First, the method depends on the initial position of the snake, which needs to be carefully initialized to be close to the region of interest to avoid being distracted by noise. Secondly, the method depends on the quality of the images, having drawbacks when applied to low-contrast images. The third difficulty is that the snake may not be able to fully adapt to the structure in some parts, whereas in other parts it may already be too flexible and leak into neighboring structures. Lastly, the method is complicated by the need to determine several parameters prior to iterations. The desired result is obtained by running the algorithm several times using different sets of parameter values until satisfactory performance is obtained.

Thus, it would be advantageous to develop a nonparametric active contour method, but to date, only a few of these approaches have been investigated [22–24]. In one such report [22], the nonparametric model translates the parameter search problem into a density estimation problem. The shortcomings of this approach include the selection of kernel size and the risk of not detecting the boundary concavities, although the issue of active contours as boundary concavities is addressed by perturbing the points on the snake after convergence. The statistical prior information about the shapes of the objects to be segmented has also been used as a solution for the parameter search problem in active contours [24]. However, the snake may not fit into low edge density regions along the object boundary or into boundary concavities, depending on the kernel size and the distance of the initialization curve from the ridges. In another report examining a nonparametric active contour approach [23], the authors proposed a statistical snake technique that is based on minimizing stochastic complexity (minimum description length principle). Although this approach can result in accurate segmentation if the distribution information is available as a priori, this is difficult in practice as images and their features are diverse.

Therefore, in the present work, we overcome the aforementioned disadvantages and difficulties associated with traditional active contour models by using an interior point search method based on a nonparametric snake model. Two key practical advantages of the proposed model are that it does not require initialization of the snake or parameter selections by the user.

The remainder of this paper is organized as follows: Section 2 illustrates the proposed nonparametric active contour scheme; Sect. 3 presents the experimental setup, results, and performance evaluation and compares the performance of the proposed model with the traditional snake model; Sect. 4 provides our conclusions and future plans.

## 2 Proposed nonparametric model

### 2.1 Mathematical model

In image processing, one objective is to identify one or more shapes in an edge map. The edge points identified in image processing techniques are usually insufficient for users to recognize the shape. Many methods have been developed and applied to identify additional edge points as discussed in Sect. 1. We develop here a new algorithm that identifies edge points using the interior point method, which has not been previously considered in the image segmentation literature. Compared with the previously developed approaches, the interior point method developed here has two key advantages. First, the interior point method starts from any randomly selected snake points, whether they are close to or far from the contour to be identified. Secondly, the interior point method requires no intervention from users. It uses information in the edge map and executes automatically until the optimal solution (snake points that form an edge or a contour of a shape) is identified. The difference between any point in an edge map, which may or may not be an edge point, and an edge point in the edge map is $R - |P_t - P_m|$, where $t$ is the index of all points in the edge map and $t = 1, \ldots, T$ with a total of $T$ points. The variable $m$ is the index of edge points and $m = 1, \ldots, M$ with a total of $M$ edge points that have already been identified in preprocessing. $M < T$. $P_t$ is the pixel value of point $t$ in the edge map. $P_m$ is the pixel value of edge point $m$. $|P_t - P_m|$ is the absolute value of the difference between $P_t$ and $P_m$, and $|P_t - P_m| \geq 0$. $R$ is the maximum absolute difference between any two points in the edge map. $R - |P_t - P_m| \geq 0$.

Similarly, the difference between a point in the edge map and a snake point is computed as $R - |P_t - P_k|$, where $k = 1, \ldots, K$, and there are a total of $K$ snake points. $P_k$ is the pixel value of snake point $k$. $|P_t - P_k|$ is the absolute value of the difference between $P_t$ and $P_k$, and $|P_t - P_k| \geq 0$. $R = \max_{k,m,t} |P_t - P_{\text{korm}}|$. $R - |P_t - P_k| \geq 0$. The difference between a point $t$ and all edge points $M$ is $R - \frac{1}{M} \sum_{m=1}^{M} |P_t - P_m|$. The difference between a point $t$ and all $K$ snake points is $R - \frac{1}{K} \sum_{k=1}^{K} |P_t - P_k|$.

The difference between all $T$ points in the edge map and all $M$ edge points is a vector $V^E$ with $T$ components (Eq. 1). Each component in $V^E$ represents the difference between a point in the edge map and all $M$ edge points. Similarly, the difference between all $T$ points in the edge map and all $K$ snake points is a vector $V^S$ with $T$ components (Eq. 2). Each component in $V^S$ represents the difference between a point in the edge map and all $K$ snake points.

$$V^E = \begin{vmatrix} R - \frac{1}{M} \sum_{m=1}^{M} |P_1 - P_m| \\ R - \frac{1}{M} \sum_{m=1}^{M} |P_2 - P_m| \\ . \\ . \\ . \\ R - \frac{1}{M} \sum_{m=1}^{M} |P_t - P_m| \end{vmatrix} \qquad (1)$$

$$V^{\mathrm{S}} = \begin{vmatrix} R - \frac{1}{K} \sum_{k=1}^{K} |P_1 - P_k| \\ R - \frac{1}{K} \sum_{k=1}^{K} |P_2 - P_k| \\ . \\ . \\ . \\ R - \frac{1}{K} \sum_{k=1}^{K} |P_t - P_k| \end{vmatrix} \quad (2)$$

In image processing, given a set of $M$ edge points identified in preprocessing, a set of $K$ snake points is selected and an algorithm is applied to move the snake points toward the edge of a shape. At the conclusion of the algorithm, the snake points form the edge of the shape. This problem may be modeled as a dot (inner) product of $V^{\mathrm{E}}$ and $V^{\mathrm{S}}$, i.e., $V^{\mathrm{E}} \cdot V^{\mathrm{S}}$. As snake points move, $V^{\mathrm{S}}$ changes. The $V^{\mathrm{S}}$ that is the same as or closest to $V^{\mathrm{E}}$ maximizes $V^{\mathrm{E}} \cdot V^{\mathrm{S}}$. Equation 3 is the mathematical model whose optimal solution identifies such a $V^{\mathrm{S}}$ and provides a set of snake points that forms the edge of the shape. The constraint $V^{\mathrm{S}} \leq V^{\mathrm{E}}$ indicates that each component of $V^{\mathrm{S}}$ must be less than or equal to the corresponding component in $V^{\mathrm{E}}$.

$$\text{Maximize } V^{\mathrm{E}} \cdot V^{\mathrm{S}} \quad \text{subject to} \quad V^{\mathrm{S}} \leq V^{\mathrm{E}} \quad (3)$$

The dot product $V^{\mathrm{E}} \cdot V^{\mathrm{S}}$ in the objective function of Eq. 3 is calculated in Eq. 4, which can be expanded in Eq. 5. There are three terms in Eq. 5, $R^2$, $\frac{R}{M} \sum_{m=1}^{M} |P_t - P_m|$, and $\frac{1}{K} \left( R - \frac{1}{M} \sum_{m=1}^{M} |P_t - P_m| \right) \sum_{k=1}^{K} |P_t - P_k|$. The first two terms are known. The third term is a function of the snake points. As the snake points change, $P_k$ changes, and the third term varies. The objective function in Eq. 3 is therefore equivalent to minimizing the summation of the third term. Equation 6 is the revised mathematical model.

$$V^{\mathrm{E}} \cdot V^{\mathrm{S}} = \sum_{t=1}^{T} \left[ \left( R - \frac{1}{M} \sum_{m=1}^{M} |P_t - P_m| \right) \left( R - \frac{1}{K} \sum_{k=1}^{K} |P_t - P_k| \right) \right] \quad (4)$$

$$V^{\mathrm{E}} \cdot V^{\mathrm{S}} = \sum_{t=1}^{T} \left[ R^2 - \frac{R}{M} \sum_{m=1}^{M} |P_t - P_m| - \frac{1}{K} \left( R - \frac{1}{M} \sum_{m=1}^{M} |P_t - P_m| \right) \sum_{k=1}^{K} |P_t - P_k| \right] \quad (5)$$

$$\text{minimize} \sum_{t=1}^{T} \left[ \frac{1}{K} \left( R - \frac{1}{M} \sum_{m=1}^{M} |P_t - P_m| \right) \sum_{k=1}^{K} |P_t - P_k| \right] \quad \text{subject to} \, V^{\mathrm{S}} \leq V^{\mathrm{E}} \quad (6)$$

Let $c_t = \frac{1}{K} \left( R - \frac{1}{M} \sum_{m=1}^{M} |P_t - P_m| \right)$; $c_t$ is known and does not change as the snake points move. The objective function in Eq. 6 is simplified as $\sum_{t=1}^{T} \left( c_t \sum_{k=1}^{K} |P_t - P_k| \right)$. The constraint in Eq. 6 is expressed as a system of total $T$ constraints (Eq. 7), which may be rewritten in Eq. 8. The right-hand side of Eq. 8 may be rewritten in Eq. 9 using $c_t$. The mathematical model in Eqs. 3 and 6 is transformed to the model in Eq. 10. This is a nonlinear programming (NLP) model with a nonlinear objective function and a total of $T$ nonlinear constraints. The decision variables in Eq. 10 are $P_k$'s, which are pixel values of the snake points. There are a total of $K$ decision variables.

$$R - \frac{1}{K} \sum_{k=1}^{K} |P_t - P_k| \leq R - \frac{1}{M} \sum_{m=1}^{M} |P_t - P_m|, \quad t = 1, \dots, T \quad (7)$$

$$\sum_{k=1}^{K} |P_t - P_k| \geq \frac{K}{M} \sum_{m=1}^{M} |P_t - P_m|, \quad t = 1, \dots, T \quad (8)$$

$$\frac{K}{M} \sum_{m=1}^{M} |P_t - P_m| = K \left( \frac{1}{M} \sum_{m=1}^{M} |P_t - P_m| \right) = K (R - K c_t) = KR - K^2 c_t \quad (9)$$

$$\text{minimize} \sum_{t=1}^{T} \left( c_t \sum_{k=1}^{K} |P_t - P_k| \right)$$
$$\text{subject to} \sum_{k=1}^{K} |P_t - P_k| \geq KR - K^2 c_t, t = 1, \dots, T$$
$$P_k \geq 0, k = 1, \dots, K \quad (10)$$

Because $c_t \geq 0$, the objective function of Eq. 10 is to minimize the summation of positively weighted absolute values. These particular types of nonlinear objective functions may be converted to linear objective functions by adding non-negative variables and a system of equality constraints [25]. Let $P_t - P_k = a_{t,k} - b_{t,k}$, $a_{t,k}, b_{t,k} \geq 0$. Equation 10 is transformed into an LP model in Eq. 11. Since Eq. 11 has a minimize objective function, for each pair of $a_{t,k}$ and $b_{t,k}$, at least one of them becomes zero at the optimal solution, i.e., $|P_t - P_k| = a_{t,k} + b_{t,k}$ at the optimal solution. The constraint $\sum_{k=1}^{K}(a_{t,k} + b_{t,k}) \geq KR - K^2 c_t$ in Eq. 9 is relaxed compared with the constraint $\sum_{k=1}^{K}|P_t - P_k| \geq KR - K^2 c_t$ in Eq. 10 because $a_{t,k} + b_{t,k} \geq |P_t - P_k|$. These two constraints are equivalent at the optimal solution because at least one of $a_{t,k}$ and $b_{t,k}$ is zero.

$$\text{minimize} \sum_{t=1}^{T}\left[c_t \sum_{k=1}^{K}(a_{t,k} + b_{t,k})\right]$$

$$\text{subject to} \quad \sum_{k=1}^{K}(a_{t,k} + b_{t,k}) \geq KR$$

$$-K^2 c_t, \; t = 1, \ldots, T$$

$$P_t - P_k = a_{t,k} - b_{t,k}, \quad k = 1, \ldots, K; \quad t = 1, \ldots, T$$

$$a_{t,k}, b_{t,k}, P_k \geq 0, \quad k = 1, \ldots, K; \quad t = 1, \ldots, T \quad (11)$$

Two methods may be used to find the optimal solution to an LP model: the simplex method and the interior point method. Before any method can be applied, however, Eq. (11) must be converted to a standard-form LP (Eq. 12). There is a total of $2KT + K + T$ variables and $KT + T$ linear constraints in Eq. (12). When the simplex method is used to find the optimal solution to Eq. (12), in the worst-case scenario, $\frac{(2KT + K + T)!}{(KT + T)!(KT + K)!}$ extreme points are visited before the optimal solution is identified. Therefore, for large LP models, the interior point method is more efficient.

$$\text{minimize} \sum_{t=1}^{T}\left[c_t \sum_{k=1}^{K}(a_{t,k} + b_{t,k})\right]$$

$$\text{subject to} \quad \sum_{k=1}^{K}(a_{t,k} + b_{t,k})$$

$$-e_t = KR - K^2 c_t, \; t = 1, \ldots, T$$

$$a_{t,k} - b_{t,k} + P_k = P_t, \quad k = 1, \ldots, K; \quad t = 1, \ldots, T$$

$$7a_{t,k}, b_{t,k}, e_t, P_k \geq 0, \quad k = 1, \ldots, K; \quad t = 1, \ldots, T \quad (12)$$

## 2.2 Interior point method

Let $X^0$ represent an initial solution to Eq. 12. $X^0$ represents a set of $K$ snake points randomly selected from an edge map. The interior point method is applied in iterations to improve $X^0$ and find the set of snake points that forms the edge of a shape. At each iteration of the interior point method, one or more snake points are replaced by other points in the edge map. As a result, the difference between snake points and points in the edge map changes, i.e., $V^S$ changes. A new solution, $X^n$, $n = 1, 2, \ldots$, is formed in each iteration, which produces a $V^S$ that is closer than the previous solution to $V^E$. The interior point method stops when further improvement for $X^n$ cannot be achieved. The optimal solution, $X^*$, is the closest to $V^E$. One or more components of $X^*$ equal zero, indicating $X^*$ becomes an extreme point (or boundary point if there are multiple optimal solutions).

Two main steps in the interior point method are (1) determining an initial solution $X^0$ and (2) iteratively improving $X^0$ to identify the optimal solution $X^*$.

### 2.2.1 Initial solution

The interior point method begins at an initial solution, $X^0$, which contains values for a set of nonnegative variables. The $X^0$ must be strictly positive and represents an initial set of snake points, which may be selected randomly from an edge map. The model in Eq. 12 has a total of $2KT + K + T$ variables and $KT + T$ linear constraints. As an example, suppose a small portion (1000 points; $T = 1000$) of a large edge map is used to identify 100 edge points ($K = 100$). It is necessary to determine values for 201,100 components in $X^0$, and these values must satisfy 101,000 constraints. Furthermore, all components in $X^0$ must be strictly positive.

The interior point method is sensitive to the initial solution [26,27]. Large positive values in the initial solution require more iterations to reach the optimal solution and therefore more computation time for the algorithm to converge. If some components in the initial solution are small and others are relatively large, the subsequent solutions may become badly scaled before the algorithm converges to the optimal solution. A systematic approach should be followed to determine values in the initial solution.

First, values for $P_k$ are determined given the initial set of snake points. Secondly, the maximum of the absolute difference between $P_k$ and $P_t$ is calculated. Let $\max_{k,t}|P_k - P_t|$ represent the maximum difference between any two points. Let $b_{t,k} = \max_{k,t}|P_k - P_t| + \varepsilon$, $\varepsilon > 0$. $b_{t,k}$ is therefore strictly positive. This guarantees that $a_{t,k}$ is strictly positive because $a_{t,k} = b_{t,k} - (P_k - P_t) = \max_{k,t}|P_k - P_t| + \varepsilon - (P_k - P_t) \geq \varepsilon$, $\varepsilon > 0$. Finally, $e_t = \sum_{k=1}^{K}(a_{t,k} + b_{t,k}) - (KR - K^2 c_t) = \sum_{k=1}^{K}[2(\max_{k,t}|P_k - P_t| + \varepsilon) - (P_k - P_t)] - (KR - K^2 c_t)$.

The value for $\varepsilon$ should be sufficiently large to ensure that $e_t > 0$. In summary, the values for $P_k$ in the initial solution $X^0$ are determined by the choice of an initial set of snake points. The values for $a_{t,k}$, $b_{t,k}$, and $e_t$ in $X^0$ are then determined in Eqs. 13–15, respectively, where $\varepsilon > 0$.

$$a_{t,k} = \max_{k,t}|P_k - P_t| + \varepsilon - (P_k - P_t) \quad (13)$$

$$b_{t,k} = \max_{k,t}|P_k - P_t| + \varepsilon \quad (14)$$

$$e_t = \sum_{k=1}^{K}[2(\max_{k,t}|P_k - P_t| + \varepsilon)$$

$$- (P_k - P_t)] - (KR - K^2 c_t) \quad (15)$$

### 2.2.2 Iterations

The initial solution, $X^0$, to Eq. 12 is a $2KT + K + T$ by 1 column vector. Once $X^0$ is determined, the interior point method uses affine scaling and applies a sequence of matrix manipulations to calculate the next solution. Let $X^n$ be a current solution to Eq. 12. The next solution, $X^{n+1}$, is calculated as follows: $X^{n+1} = X^n + \lambda_n \cdot \Delta_n$, where $\Delta_n$ is an improving search direction and a $2KT + K + T$ by 1 column vector. $\lambda_n$ is a step size, indicating how much $X^n$ should move along $\Delta_n$ to reach $X^{n+1}$. $\lambda_n$ is a scalar, and $\lambda_n > 0$. Let $A$ be the coefficient matrix for constraints in Eq. 12. $A$ is a $KT + T$ by $2KT + K + T$ matrix. Let $C$ be the coefficient vector for the objective function in Eq. 12. $C$ is a 1 by $2KT + K + T$ row vector. Let $X_n$ be the diagonal matrix of the current solution $X^n$. Thus, $\Delta_n$ and $\lambda_n$ are calculated in Eqs. 16 and 17, respectively.

$$\Delta_n = -X_n \left\{ I - (A \times X_n)' \right.$$
$$\left. \left[ (A \times X_n)(A \times X_n)' \right]^{-1} (A \times X_n) \right\} (C \times X_n)' \tag{16}$$

$$\lambda_n = \frac{1}{X_n^{-1} \times \Delta_n} \tag{17}$$

## 3 Experimental setup and results

This section presents the evaluation of the proposed nonparametric snake model. A representative set of synthetic images is considered in the first step of the evaluation. Two real images, an airplane and a magnetic resonance (MR) images, are also tested to evaluate the model. The evaluation demonstrates that the proposed model provides good results and is a simple yet robust segmentation technique that avoids all the aforementioned complications of previously developed active contour models and nonparametric snake models.

### 3.1 Dataset description

The representative set of images was used for evaluation. The first set of binary images was generated using MATLAB 2013a, and the size of the images was $400 \times 320$ pixels. The shapes of these images (arrow, heart, rectangle, star, and a combination of arrow, rectangle, and star) were selected to represent different shape features, such as sharp corners, concavity, and convexity. An additional set using one of these images, Fig. 1a, with three levels of noise (Gaussian noise with a mean of zero and standard deviations of 25, 50, and 75 shown in Fig. 2a) was added to the representative set to test how well the proposed model deals with noisy images. The representative set also included a color photograph of an airplane, two grayscale MR images, and a computed tomog-raphy (CT) image, Fig. 3a. Different formats and sizes were used to test the robustness of the proposed model. The test images for the nonparametric snake model-based segmentation were loaded into MATLAB 2013a on a workstation with a 3.20 GHz Core i5-6500 CPU processor and 16 GB of RAM.

### 3.2 Binary and continuous edge maps

Binary and continuous edge maps are required to test the proposed nonparametric snake model. The binary edge map provides apriori information about the locations of the objects that are segmented (i.e., a limited number of edge points identified in preprocessing), whereas the continuous edge map provides numerical values of the pixels. The binary edge map was built using an enhanced edge detection algorithm that we recently developed [28]. A continuous edge map is derived by finding gradient magnitudes of the images using $G(p) = \sqrt{G(p)_x^2 + G(p)_y^2}$, with $G(p)_x^2$ $x$-derivative and $G(p)_y^2$ $y$-derivative. The values are then normalized by dividing the values by the maximum value. Therefore, all gradient values are between 0 and 1.

The proposed model was tested and validated on the set of representative images shown in Fig. 1. Figure 1a shows the originally generated binary images, while (b) and (c) provide binary edge and continuous edge maps, respectively, and (d) indicates the contours of the images shown in (a) as generated by the proposed model. The results indicate that based on comparing the positions of the generated pixels with a truth table that expresses the true positions of manually found object boundaries the proposed nonparametric snake model performs well on these images. It is especially noteworthy that the proposed model found the contour of the image composed of multiple objects (star, arrow, and rectangle; Fig. 1, bottom row) without any human intervention to initialize either multiple snakes or one big snake (size of entire image).

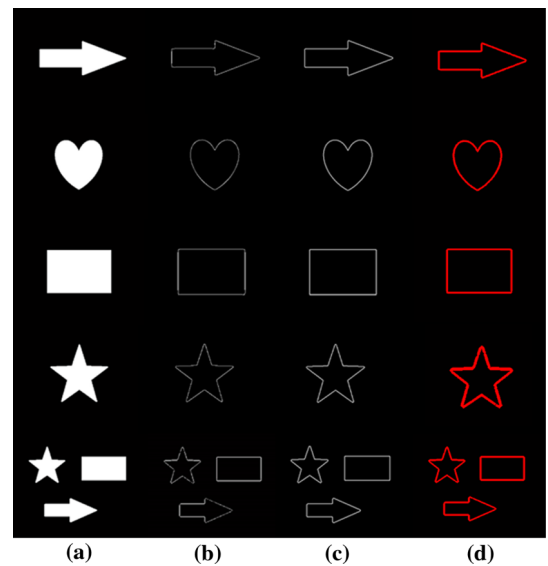

**Fig. 1** Validation of the proposed nonparametric snake model using a representative set of synthetic images. **a** Originally generated binary images (from Fig. 1a, e–h). **b** Binary edge maps. **c** Continuous edge maps. **d** Contours generated by the proposed model

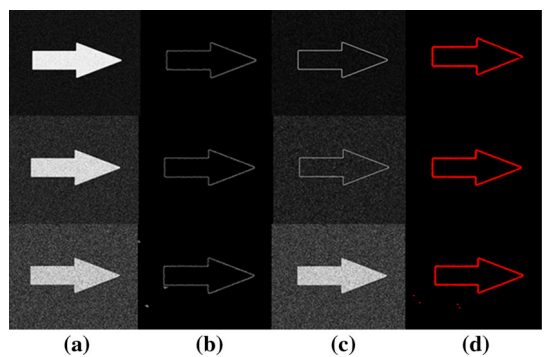

**Fig. 2** Validation of the proposed nonparametric snake model using a single image with different levels of Gaussian noise (from Fig. 1b–d). Gaussian noise with a mean of zero and standard deviations of 25 (*top row*), 50 (*middle row*), and 75 (*bottom row*). **a** Originally generated binary images. **b** Binary edge maps. **c** Continuous edge maps. **d** Contours generated by the proposed model

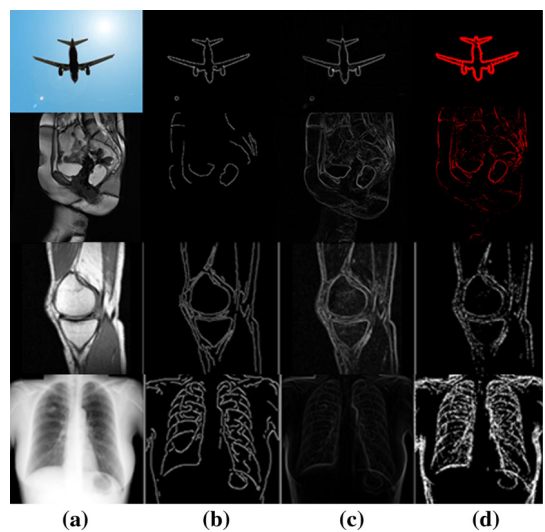

**Fig. 3** Validation of the proposed nonparametric snake model using representative real images. **a** From *top* to *bottom*: airplane, MRI, knee, and lungs images with a *grayscale* range. **b** Binary edge maps. **c** Continuous edge maps. **d** Contours generated by the proposed model

**Table 1** Performance comparison of the proposed model with the traditional snake

|  | DSI of the proposed model | DSI of traditional active contour |
|---|---|---|
| Arrow | 0.990 | 0.988 |
| Heart | 0.993 | 0.987 |
| Rectangle | 0.993 | 0.994 |
| Star | 0.995 | 0.985 |
| Multiple | 0.991 | 0.986 |
| Arrow-noise1 | 0.996 | 0.994 |
| Arrow-noise2 | 0.992 | 0.942 |
| Arrow-noise3 | 0.996 | 0.617 |
| Airplane | 0.989 | 0.974 |
| MRI | 0.928 | 0.424 |
| Knee | 0.962 | 0.547 |
| Lungs | 0.912 | 0.436 |

In Fig. 2, we show the performance of the proposed nonparametric model on noisy synthetic images using the shape of the arrow and different levels of noise. The model identifies the contour of these arrows without being affected by the noise level. This example indicates that the proposed model overcomes a major disadvantage of the traditional snake model and is robust in the presence of noise.

After presenting the validation using synthetic images and images with different levels of noise, we now present the results of applying the proposed model to more challenging images: a photograph consisting of $400 \times 320$ pixels of an airplane and an MR image of $512 \times 512$ pixels with a grayscale range. For both examples, we selected random snake points over the images because the proposed nonparametric snake has no capture range problems. Thus, the proposed model requires no preprocessing steps, such as filtering or enhancing. Figure 3d presents the results provided by the proposed model, showing clearly defined borders of the airplane and the anatomical structures, although a few edge points in the latter are missing. The missing edge problem can be overcome by applying the model only to the region of interest.

The performance of the proposed segmentation method was measured by quantifying the region overlap between the manual and automated segmentations using the dice similarity index (DSI).

The performance of the proposed segmentation model and traditional active contour is presented in Table 1. It can be observed that the proposed method can correctly detect all objects with higher accuracy, (DSI > 0.95). On the other hand, the traditional active contour model can only accurately detect the objects other than "arrow-noise 2," "arrow-noise 3," "MRI," "knee," and "lungs" because of the noise level and complexity. Compared to the traditional active contours, the proposed segmentation method achieved the highest DSI for all the objects except for "rectangle," where it performed very similar to the traditional active contour model.

In terms of computational time, the traditional snake provides faster than the proposed model if the optimal parameters can be used. To do that, the trial-and-error approach which is very time-consuming to reach the better segmentation results has been used. The proposed model is computationally more expensive than the traditional snake. However, this can be solved using parallel computing. For the traditional snake, segmentation time of the complex images such as MRI, knee, and lungs increases significantly as it requires exhaustive search of parameters for the optimization process.

**Table 2** Pairwise comparison of two methods

| | Difference between the areas | Standard error | 95% confidence interval | $p$ value |
|---|---|---|---|---|
| Arrow | 0.0102 | 0.0680 | −0.009 to 0.021 | $p = 0.720$ |
| Heart | 0.0144 | 0.0814 | −0.014 to 0.091 | $p = 0.938$ |
| Rectangle | 0.0032 | 0.0694 | −0.013 to 0.232 | $p = 0.612$ |
| Star | 0.0109 | 0.0611 | −0.017 to 0.521 | $p = 0.620$ |
| Multiple | 0.0228 | 0.0681 | −0.013 to 0.098 | $p = 0.568$ |
| Arrow-noise1 | 0.0166 | 0.0847 | −0.015 to 0.103 | $p = 0.816$ |
| Arrow-noise2 | 0.0158 | 0.0382 | 0.165 to 0.369 | $p = 0.054$ |
| Arrow-noise3 | 0.0176 | 0.0228 | 0.008 to 0.425 | $p = 0.022$ |
| Airplane | 0.0066 | 0.0791 | −0.005 to 0.012 | $p = 0.717$ |
| MRI | 0.0278 | 0.0409 | 0.010 to 0.925 | $p = 0.016$ |
| Knee | 0.0248 | 0.0484 | 0.010 to 0.436 | $p = 0.021$ |
| Lungs | 0.0486 | 0.0350 | 0.043 to 0.198 | $p = 0.019$ |

### 3.3 Comparison of the proposed and traditional snake models

Finally, we compared the performance of the proposed nonparametric snake model with a traditional snake model using the image composed of multiple objects and the image of an arrow with three levels of noise. As discussed previously, one of the drawbacks of the traditional active contour is its high level of sensitivity to the control parameters. For our experiments, the trial-and-error approach has been used for the adjustment of the parameters, which makes the process very time-consuming, and frequently it cannot guarantee that optimal values are obtained. The traditional snake model identified the borders of the multiple objects and the arrows with the two lower levels of noise when it was provided with properly chosen parameters. However, this is the case only when the aforementioned suitable parameter values are known or can be derived from the image in the preprocessing stage. By contrast, the traditional snake model failed to find only the border of the object in the image with the highest level of noise level (Gaussian noise with mean of zero and standard deviation of 75). It also found several other points, which are noise.

A statistical significant test using DeLong's test [29] has also been conducted to assess the difference between the proposed method and the traditional snake. The result of pairwise comparisons is presented in Table 2: the difference between the areas, the standard error, the 95% confidence interval for the difference, and $p$ value. If $p$ is less than the conventional 5% ($p < 0.05$), the conclusion is that two methods are significantly different for the object identified. It can be observed that these results also support the performance evaluations presented in Table 1. The identified regions by the proposed model and the active contour are significantly different in "arrow-noise 2," "arrow- noise 3," "MRI," "knee," and "lungs".

## 4 Conclusions

Active contour or snake models are useful algorithms for image segmentation in the fields of image analysis and computer-aided diagnosis. However, difficulties in parameter selection limit their use in high-throughput scenarios or by non-experts. Although many different models for active contours have been reported in the literature, all of them require properly selected parameters to achieve satisfactory results. In this study, we exploited underlying binary and continuous edge information in edge maps to develop an alternative nonparametric model. Our validation trials of the proposed model indicate that it provides more robust results than a traditional parametric snake model, in which users must define global parameters. Thus, the first key advantage of the proposed nonparametric snake model is its robustness. The second important advantage is that the proposed model requires no user interaction for initializing the snakes before running the algorithm; it requires only edge maps as input. In addition to these advantages, the proposed model performs well with low-contrast images, such as MR images, which not only require a high number of iterations by the traditional snake model but also result in unfaithful segmented structures. A final key advantage of the proposed model over traditional snake models is that the noise level in an image is not an issue; thus, no preprocessing is required to find the contour in a noisy image. The proposed nonparametric, interior point method can be applied to three-dimensional (3-D) images to identify edges of 3-D shapes. However, even though the proposed model is capable to segment 3-D objects, it is computationally expensive. Thus, we are currently working on improving the proposed method to reduce the computation time for 3-D objects. The proposed model can also identify edges of non-continuous shapes. For instance, the image in the last row of Fig. 1 consists of three shapes that are disconnected from each other. The proposed model was able to successfully identify all three shapes in the image. In 3-D images, occlusion may occur if a 3-D shape blocks other shapes from view. The interior point method cannot identify the edges of shapes that are blocked from view because pixel values of these shapes are not available. Another weakness of the interior point method is that it requires a substantial amount of computation time. The computation time varies and depends on the size of an image, the number of edge points to be identified, and the performance of the computer. One approach to overcome this weakness is parallel computing using multiple computers. The interior point method can be implemented on multiple computers at the same time to identify edge points. In our future work, we aim to extend this analysis to a nonlinear formulation of the problem. This will eliminate using several constraints considered in the current proposed model. It is not expected to improve the present results because the model presented here is already robust; however, it may reduce computation time, producing the same good results in less time than the current proposed model.

## References

1. Amza, C.G., Cicic, D.T.: Industrial image processing using fuzzy-logic. Procedia Eng. **100**, 492–498 (2015)
2. Islam, M.J., Ahmadi, M., Sid-Ahmed, M.A.: Image processing techniques for quality inspection of gelatin capsules in pharmaceutical applications. In: 2008 10th International Conference on Control, Automation, Robotics and Vision. (2008)
3. Shastry, S., et al.: A novel algorithm for optical character recognition (OCR). In: 2013 International Mutli-Conference on Automation, Computing, Communication, Control and Compressed Sensing (iMac4s).(2013)
4. Ojha, S., Sakhare, S.: Image processing techniques for object tracking in video surveillance- A survey . In: 2015 International Conference on Pervasive Computing (ICPC). (2015)
5. Onal, S., et al.: Fully automated localization of multiple pelvic bone structures on MRI. Conf. Proc. IEEE Eng. Med. Biol. Soc. **2014**, 3353–6 (2014)
6. Onal, S., et al.: Automated localization of multiple pelvic bone structures on MRI. IEEE J. Biomed. Health Inform. **20**(1), 249–55 (2016)
7. Mao, J., Jain, A.: Texture classification and segmentation using multiresolution simultaneous autoregressive models. Pattern Recogn. **25**, 173–188 (1992)
8. Hofmann, T., Puzicha, J., Buhmann, J.: Unsupervised texture segmentation in a deterministic annealing framework. IEEE Trans. Pattern Anal. Mach. Intell. **20**(8), 803–818 (1998)
9. Hall, L.O., Bensaid, A.M., Clarke, L.P., Velthuizen, R.P., Silbiger, M.S., Bezdek, J.C.: A comparison of neural network and fuzzy clustering techniques in segmenting magnetic resonance images of the brain. IEEE Trans. Neural Netw. **3**(5), 672–681 (1992)
10. Tremeau, A., Borel, N.: A region growing and merging algorithm to colour segmentation. Pattern Recogn. **30**(7), 1191–1203 (1997)
11. Hojjatoleslami, S., Kittler, J.: Region growing: A new approach. IEEE Trans. Image Process. **7**, 1079–1084 (1998)
12. Bao, P., Zhang, L.: Noise reduction for magnetic resonance images via adaptive multiscale products thresholding. IEEE Trans. Med. Imaging **22**(9), 1089–1099 (2003)
13. Bao, P., Zhang, L., Wu, X.: Canny edge detection enhancement by scale multiplication. IEEE Trans. Pattern Anal. Mach. Intell. **27**(9), 1485–1490 (2005)
14. Canny, J.: A computational approach to edge detection. IEEE Trans. Pattern Anal. Mach. Intell. **8**(6), 679–698 (1986)
15. Kass, M., Witkin, A., Terzopoulos, D.: Snakes: Active contour models. Int. J. Comput. Vis. **1**(4), 321–331 (1988)
16. Terzopoulos, D., Fleischer, K.: Deformable models. Visual Comput. **4**(6), 306–331 (1988)
17. Cootes, T.F., Cooper, D.: Active shape models—their training and application. Comput. Vis. Image Underst. **6**(1), 38–59 (1995)
18. Chan, T.F., Vese, L.A.: Active contours without edges. IEEE Trans. Image Process. **10**(2), 266–77 (2001)
19. Allili, M.S., Ziou, D.: Active contours for video object tracking using region, boundary and shape information. Signal Image Video Process. **1**(2), 101–117 (2007)
20. Ge, Q., et al.: Active contour evolved by joint probability classification on Riemannian manifold. Signal Image and Video Process. **10**(7), 1257–1264 (2016)
21. Reska, D., Boldak, C., Kretowski, M.: Towards multi-stage texture-based active contour image segmentation. Signal Image and Video Process. **15**(5), 809–816 (2017)
22. Ozertem, U., Erdogmus, D.: Nonparametric snakes. IEEE Trans. Image Process. **16**(9), 2361–8 (2007)
23. Martin, P., et al.: Nonparametric statistical snake based on the minimum stochastic complexity. IEEE Trans. Image Process. **15**(9), 2762–70 (2006)
24. Kim, J., Çetin, M., Willsky, A.S.: Nonparametric shape priors for active contour-based image segmentation. Signal Process. **87**(12), 3021–3044 (2007)
25. Rardin, L.R.: Optimization in Operations Research, 2nd edn. Prentice Hall, Upper Saddle (2017)
26. Simon, J.-W., Höwer, D., Weichert, D.: A starting-point strategy for interior-point algorithms for shakedown analysis of engineering structures. Eng. Optim. **46**(5), 648–668 (2014)
27. Gondzio, J.: Crash start of interior point methods. Eur. J. Oper. Res. **255**(1), 308–314 (2016)
28. Onal, S., et al.: Automatic vertebra segmentation on dynamic magnetic resonance imaging. J. Med. Imaging **4**(1), 014504–014504 (2017)
29. DeLong, E.R., DeLong, D.M., Clarke-Pearson, D.L.: Comparing the areas under two or more correlated receiver operating characteristic curves: A nonparametric approach. Biometrics **44**(3), 837–845 (1988)